\documentclass[shortnote,twocolumn]{jpsj3}
\usepackage{txfonts}
\usepackage{braket}
\usepackage{dcolumn}
\usepackage{bm}
\usepackage{color}

\title{Phase diagram of the Kitaev-Heisenberg model using various finite-size clusters}

\author{Masahiro Kadosawa$^1$\thanks{afna1728@chiba-u.jp}, Masaaki Nakamura$^2$, Yukinori Ohta$^1$, and Satoshi Nishimoto$^{3,4}$}
\inst{$^1$Department of Physics, Chiba University, Chiba 263-8522, Japan \\
$^2$Department of Physics, Ehime University, Ehime 790-8577, Japan \\
$^3$Department of Physics, Technical University Dresden, 01069 Dresden, Germany \\ 
$^4$Institute for Theoretical Solid State Physics, IFW Dresden, 01171 Dresden, Germany
} 

\abst{We estimate phase boundaries of four ordered and two spin-liquid phases
for the spin-$\frac{1}{2}$ Kitaev-Heisenberg (KH) model using four kinds of
relatively-small clusters, based on the second derivative of ground-state
energy. The estimated values are compared between the clusters as well as the
previous iPEPS results. We thus find that the boundaries can be accurately
estimated within limited-size clusters. The used clusters may appropriate to
study higher-$S$ KH models having less fluctuations.}

\begin{document}
\maketitle

The Kitaev model~\cite{Kitaev2006} consists of Ising bond-direction
dependent interactions on a honeycomb lattice. The ground state is
exactly known to be a quantum spin liquid, the so-called ``Kitaev
spin liquid (KSL)''. Since the microscopic origin of such Kitaev-type
interactions in the $d^5$ transition metal compounds with a strong
spin-orbit coupling was worked out~\cite{Jackeli2009}, there has been
a growing number of researches on the Kitaev materials~\cite{Trebst2022}.
Nonetheless, a more realistic spin model to describe the magnetic
properties of real materials is the Kitaev-Heisenberg (KH) model that
accounts for the residual Heisenberg-type couplings, although further neighbor
interactions and/or off-diagonal $\Gamma$ interactions may play important
roles in some materials. Initially, an exact-diagonalization study using
24-site periodic cluster (24PBC) claimed that the ground-state phase diagram
of the spin-$\frac{1}{2}$ KH model is composed of four magnetically ordered
phases, namely N\'eel, zigzag, ferromagnetic, and stripy as well as two KSL
phases, depending on the ratio between Kitaev and Heisenberg
interactions~\cite{Chaloupka2010}. A small size-dependence of
their critical points has been also confirmed by large-scale
simulations~\cite{Jiang2011,Iregui2014}. In resent years, 
researches on higher-$S$ Kitaev materials are increasing to seek for
new physics. However, when we study the corresponding spin models
numerically, the cluster size would be strongly limited. Thus, it is useful
to verify how accurately the phase boundaries are evaluated with limited-size
clusters for the spin-$\frac{1}{2}$ KH model where quantum fluctuations are
largest. 

In this short note, we estimate the phase boundaries with four kinds of
finite-size clusters using exact diagonalization and density-matrix
renormalization group (DMRG) methods. By comparing them with referring to
the previous iPEPS results~\cite{Iregui2014}, we suggest a few clusters
suitable for study of the higher-$S$ KH models. The Hamiltonian of
the KH model reads as
\begin{align}
	\mathcal{H}=\sum_{\langle ij \rangle\gamma} 2K^\gamma S_i^\gamma S_j^\gamma + J \sum_{\langle i,j \rangle} \mathbf{S}_i \cdot \mathbf{S}_j
	\label{model}
\end{align}
where $K^\gamma$ is the Kitaev interaction of nearest neighbor spins on three different bonds $\gamma=x,y,z$ and $J_{ij}$ is the nearest-neighbor Heisenberg
interaction. Assuming $K^x=K^y=K^z=K$, we parameterize the interactions as
$K=\sin \varphi$ and $J=\cos \varphi$ ($\sqrt{K^2+J^2}=1$ is the energy unit).
The ground state is controlled by $\varphi$. The critical values are estimated
from the peak positions in the second derivative of the ground-state energy $E$
as a function of $\varphi$, i.e., $-\frac{\partial^2E}{\partial \varphi^2}$.

\begin{figure}[tb]
	\centering
	\includegraphics[width=0.8\linewidth]{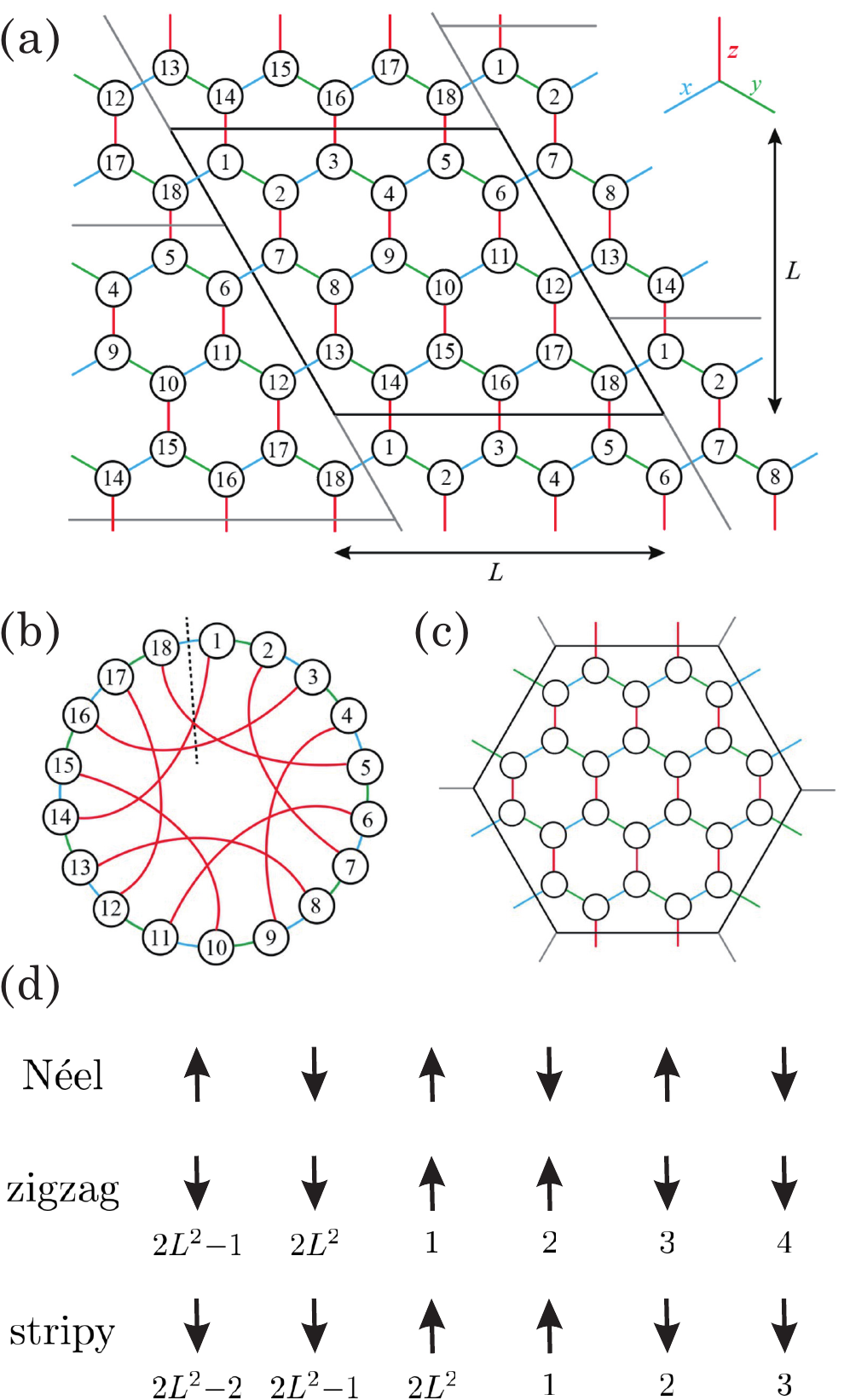}
	\caption{(a) 2D honeycomb lattice with $L = 3$, where the
		region framed by black line is the original cluster.
		(b) 1D representation of (a) using SBC. An open chain
		is created by cutting $L+1$ bonds between two sites (dotted line).
		(c) Spatially-isotropic hexagonal-shaped 24-site cluster. 
		The periodic and open clusters are denoted as 24PBC and 24OBC,
		respectively.
		(d) 1D representation of magnetically ordered states for
		the	KH model.	
	}
	\label{fig:SBC}
\end{figure}

Four kinds of clusters are employed here. First, the results for 24PBC cluster
[Fig.~\ref{fig:SBC}(c)] are used as reference data. The second is 24-site
cluster with open boundary conditions (24OBC), which is a hexagonal-shaped
cluster as 24PBC but outer sites are not connected to those in the opposite
side. Ideally, it is deemed desirable to use a spatially-isotropic cluster.
To cite a case, if a spatially-anisotropic cluster is used, the estimation
of critical points could be biased due to the lift of three-fold degeneracy
in the zigzag and stripy states~\cite{Yadav2022}. Also, it is not easy to
set up such a cluster consistent with all the ordered states. Regarding
this point, the 24PBC and 24OBC clusters are preferable.

Nevertheless, when we need to use a spatially-anisotropic cluster, it would
be good to apply spiral boundary conditions (SBC). The SBC enable us to
flexibly control the periodicity of cluster~\cite{Kadosawa2023}.
As illustrated in Fig.~\ref{fig:SBC}(a,b) a 2D honeycomb lattice with
$L\times L$ unit cells can be mapped onto a periodic chain with nearest- and
($2L-1$)th-neighbor couplings by applying SBC. 
We note that the SBC-projected chain can be also represented
as a cylinder where finite-size effects due to the short circumference are
minimized~\cite{Nakamura2021}.
We now employ an SBC-projected 32-site chain ($L=4$), which is consistent
with all the magnetically ordered states seen in the KH model. The structural
unit cell of the mapped 1D chain contains two spins as in the original 2D
lattice. The 1D representations of magnetic structures of N\'eel, zigzag,
and stripy states are illustrated in Fig.~\ref{fig:SBC}(d). The N\'eel state
possesses a staggered arrangement of spins. From a topological point of view,
it is interesting that both of the zigzag and stripy states have a two-site
periodicity but their phases are different just by one site. In the Kitaev
limits $\varphi=\frac{\pi}{2}$ and $\varphi=\frac{3}{2}\pi$, the KSL feature
can be reproduced. Namely, the spin-spin correlations are finite only between
Ising-coupled spins and zero for the others. We study periodic (32SBC-P) and open (32SBC-O) chains.

\begin{figure}[tb]
	\centering
	\includegraphics[width=0.9\linewidth]{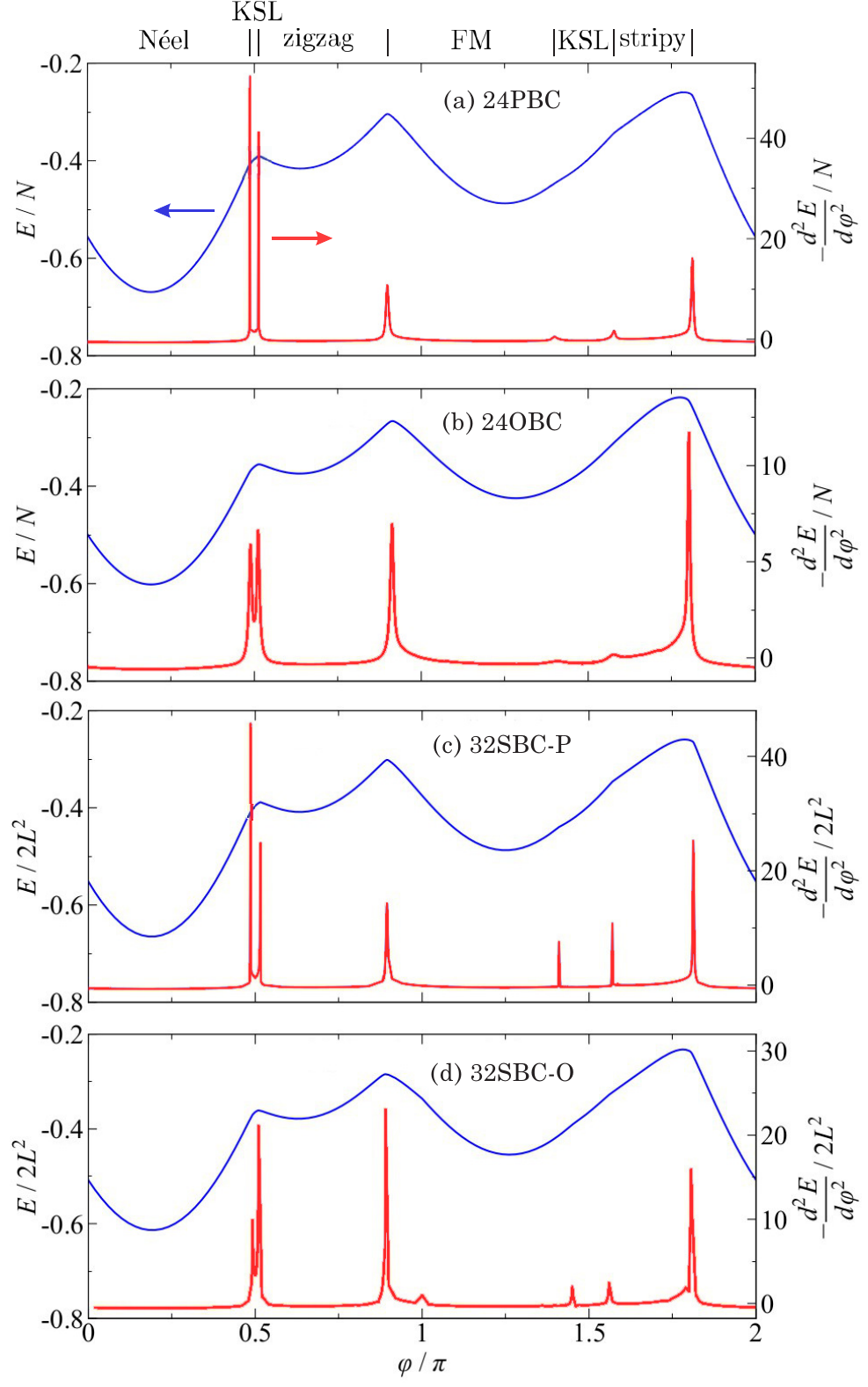}
	\caption{Ground-state energy and its second derivative as a function of
		 $\varphi$, where the phase boundaries are indicated by peaks in
		 the second derivative. The state of each period is denoted on the top.
	}
	\label{fig:energy}
\end{figure}

In Fig.~\ref{fig:energy} the ground-state energy and its second derivative
are plotted as a function of $\varphi$. For all of the four clusters,
we find that the energy behaves similarly and six peaks appear at similar
positions in the second derivative. [For 32SBC-O cluster, an additional peak
appears at $\varphi=\pi$ (explained below).] We can thus identify four ordered
and two KSL phases. It is worth noting that the second-derivative peaks
indicating the boundaries between FM KSL and its neighboring phases are
even more obvious in the use of SBC clusters.

\begin{table}[tb]
	\caption{Phase boundaries for the KH model, parameterized
		by the angle $\varphi$ (in units of $\pi$), using various clusters.
		The iPEPS results~\cite{Iregui2014} are also shown.}
	\begin{center}
		\begin{tabular}{wc{1.4cm}wc{0.8cm}wc{0.8cm}wc{1.05cm}wc{1.05cm}wc{0.9cm}}
			\hline
			\hline
			boundary & 24PBC & 24OBC & 32SBC-P & 32SBC-O & iPEPS \\
			\hline
			N\'eel/KSL   &  0.489  &  0.487  &  0.486  &  0.491 & 0.489 \\
			KSL/zigzag   &  0.511  &  0.513  &  0.516  &  0.510 & 0.511 \\
			zigzag/FM    &  0.913  &  0.898  &  0.896  &  0.889 & 0.894 \\
			FM/KSL       &  1.407  &  1.399  &  1.410  &  1.449 & 1.433 \\
			KSL/stripy   &  1.577  &  1.577  &  1.570  &  1.552 & 1.556 \\
			stripy/N\'eel&  1.802  &  1.812  &  1.812  &  1.805 & 1.817 \\
			\hline
			\hline
		\end{tabular}
	\end{center}
	\label{critialphi}
\end{table}

The estimated critical $\varphi$ values for each of the clusters are
summarized in Table~\ref{critialphi}. For comparison, the results obtained
by iPEPS~\cite{Iregui2014} are also listed. Overall, the critical values for
24OBC and 32SBC-P clusters agree to those for 24PBC cluster very well.
A largest deviation is found in the zigzag and FM boundary;
$\varphi_{\rm zigzag-FM}=0.913\pi$ for 24PBC cluster and 
$\varphi_{\rm zigzag-FM}=0.896\pi$ for 32SBC-P cluster, which corresponds to
an error of $\sim17\%$ in $K/J$. However, this $\varphi_{\rm zigzag-FM}$
value for 32SBC-P is rather closer to that obtained by iPEPS, i.e.,
$\varphi_{\rm zigzag-FM}=0.894\pi$. Possibly, since 24OBC and 32SBC-P
clusters do not have short periodic loops of bonds, their finite-size effects
may be smaller than those of 24PBC in some cases. The critical values
for 32SBC-O cluster are also in good agreement with those for 24PBC cluster.
However, the region of FM KSL seems to be a little underestimated. This is
because a spin-rotation anisotropy is induced by cutting 1 $x$-bond (or
$y$-bond) and 4 $z$-bonds when OBC is applied [see Fig.~\ref{fig:SBC}(b)].
This effect is also revealed as a second-derivative peak at $\varphi=\pi$,
where the polarization direction changes between $xy$-plane and $z$-axis.

In summary, we examined the quantitativeness of six critical values
for the spin-$\frac{1}{2}$ KH model using four kinds of clusters.
We confirmed that the phase boundaries of commensurate orders and KSL
can be accurately estimated even within limited-size clusters, based on
the second derivative of ground-state energy. Since a certain level of
accuracy for the energy is required to perform this analysis, cluster size
may be strongly limited when a larger-$S$ KH model is studied by DMRG.
Practically, either 24OBC or 32SBC-P clusters would be a good choice
for that purpose.

\begin{acknowledgment}
We thank Ulrike Nitzsche for technical support.
This work was supported by the SFB 1143 of the Deutsche Forschungsgemeinschaft and by 
Grants-in-Aid for Scientific Research from JSPS (Projects No. JP20H01849, No. JP20K03769, and No. JP21J20604).
M. K. acknowledges support from the JSPS Research Fellowship for Young Scientists.

\end{acknowledgment}

\end{document}